# Interpretation of the Klein-Gordon Probability Density

Roderick I. Sutherland

Centre for Time, University of Sydney, NSW 2006 Australia

rod.sutherland@sydney.edu.au

The fact that the "probability density" expression provided by the Klein-Gordon equation can take on negative values is usually seen as an obstacle to formulating a particle interpretation of quantum mechanics. Nevertheless, reconciling this expression with a particle ontology is quite possible once a careful distinction is drawn between the outcomes of measurements and the positions of particles between measurements. Following this path, however, points to the involvement of retrocausality, as proposed by various authors in other contexts.

## 1. Introduction

This paper is concerned with interpretations of quantum mechanics which assume that the underlying reality between measurements involves particles having definite trajectories. The most prominent example of such an interpretation is the de Broglie-Bohm model [1-3]. The paper focuses on the question of whether a suitable 4-current density expression to support such a picture is provided by each wave equation. The four most familiar wave equations for quantum mechanics are the Schrodinger, Pauli, Dirac and Klein-Gordon equations. A particle interpretation of quantum mechanics requires the existence of both a positive probability density and a conserved current density for each of these cases. Appropriate expressions are indeed provided by the standard formalism in the cases of the first three wave equations listed above, but not for the fourth. In the case of the Klein-Gordon equation, the "probability density" expression provided by the formalism is not positive definite[1]. The relevant expression for the 4-current density $j^\alpha$ has the form:

$$j^\alpha(x) = \frac{i}{2m}\psi^*(x)\,\overset{\leftrightarrow}{\partial^\alpha}\,\psi(x) \qquad (\alpha = 0,1,2,3) \qquad (1)$$

where, as usual, $\psi(x)$ is the wavefunction as a function of position $\mathbf{x}$ at time t and m is the mass of the particle in question. The units have been chosen such that $\hbar = c = 1$ and the following notation has been used: $x \equiv (t, \mathbf{x})$, $\partial^\alpha \equiv \frac{\partial}{\partial x_\alpha}$ and $\overset{\leftrightarrow}{\partial^\alpha} \equiv \overset{\rightarrow}{\partial^\alpha} - \overset{\leftarrow}{\partial^\alpha}$.

[1] This issue has been examined by many authors, e.g., Nikolic [4], Secs. 7 and 8. For Bohmian attempts to make sense of the Klein-Gordon equation, see Holland [5], Sec.12.1 and Tumulka [6], Secs. 7.2.1 and 7.3.8.

In preparation for the discussion in Sec. 5, it will be more appropriate to work in Dirac notation here, in which case the equivalent expression to Eq. (1) for the 4-current density given an initial state $|i\rangle$ is[2]:

$$j^{\alpha}\left(x|i\right)=\frac{i}{2m}\langle i|x\rangle\overset{\leftrightarrow}{\partial^{\alpha}}\langle x|i\rangle \qquad (\alpha=0,1,2,3) \qquad (2)$$

The tentative probability expression is then the time (or zeroth) component of this 4-current density:

$$j^{0}\left(x|i\right)=\frac{i}{2m}\langle i|x\rangle\overset{\leftrightarrow}{\partial^{0}}\langle x|i\rangle \qquad (3)$$

which is easily seen to take on negative values[3]. This occurs even when only positive energy states are used (see e.g., [5], Sec.12.1).

The aim here is to show that there is no impediment to a particle interpretation of the Klein-Gordon equation once a small extension to the usual formalism is included. A similar extension has already been employed in previous work [7-9], including the weak value formalism [10]. It has also been found to be needed in order to accommodate Lorentz covariance in the many-particle case [11]. As a preliminary step, the next two sections will review some relevant aspects of the standard formalism of quantum mechanics.

## 2. Standard formalism for observables

Quantum mechanics defines the possible observable quantities in terms of Hilbert space vectors. An observable is equated with a complete orthonormal set of basis vectors in such a space. Consider an observable which has a continuous spectrum of eigenvalues. Assuming an initial state $|i\rangle$, the probability density corresponding to finding a particular final eigenstate $|f\rangle$ is given by:

$$\rho\left(f\right)=\left|\langle f|i\rangle\right|^{2} \qquad (4)$$

This expression is obviously always positive.

It is well known that a problem arises in the relativistic case when an attempt is made to include position $\mathbf{x}$ in this framework. The complication is that the position eigenstates are found to be not orthogonal (see Appendix) and so position does not qualify as an observable under the above definition[4]. Note that in this situation an eigenstate $|\mathbf{x}\rangle$ will have a non-zero amplitude onto another eigenstate $|\mathbf{x}'\rangle$. This implies the surprising conclusion that if a particle is in position state $|\mathbf{x}\rangle$ at a particular time, there is a nonzero amplitude for it to be in the position state $|\mathbf{x}'\rangle$ as well at that time.

---

[2] Note that the letter i is playing two roles here, representing both “initial” and $\sqrt{-1}$.

[3] Note that it is not possible simply to take the probability density to be $j^{0}=\left|\langle x|i\rangle\right|^{2}$ because, as will be discussed later, this would not be consistent with conservation of probability.

[4] In textbooks, the more usual way to highlight that there is a problem here is by pointing out that the position operator $\mathbf{x}$ is not hermitian in the relativistic case and so cannot correspond to an observable.

Three relevant points will be mentioned here. First, the Klein-Gordon formalism does not provide any locally conserved current which is compatible with the positive expression $|\langle \mathrm{x}|\mathrm{i}\rangle|^2$ for probability density, i.e., none which has $|\langle \mathrm{x}|\mathrm{i}\rangle|^2$ as its zeroth component. Second, it is not viable in any case to use the expression $|\langle \mathrm{x}|\mathrm{i}\rangle|^2$ for position probability since the eigenstates are not orthogonal. Third, the $\mathrm{j}^0$ expression in Eq. (3) is not needed to describe the probabilities of observable results since position $\mathbf{x}$ is not an observable[5].

## 3. The position observable of Newton and Wigner

As explained in various textbooks[6], Newton and Wigner [13] have rectified the absence of a position observable by introducing an alternative observable which is able to play a similar role. This observable will be represented here by the letter $\mathbf{q}$. The basis vectors for the $\mathbf{q}$ observable are orthogonal, but do not describe fully localised positions. The probability density corresponding to a particular eigenstate $|\mathbf{q}\rangle$ is given by:

$$\rho(\mathbf{q}) = |\langle \mathbf{q}|\mathrm{i}\rangle|^2 \tag{5}$$

which is positive, as required. The $\mathbf{q}$ states are, however, not Lorentz covariant.

At this point it is important to distinguish between (i) the outcome of an actual measurement, such as a measurement of $\mathbf{q}$, and (ii) the position of a particle at times between measurements. The standard formalism is fully satisfactory for describing measurements of observables, but does not seem helpful towards constructing a particle model for times between measurements in the Klein-Gordon case. In particular, although the $\mathbf{q}$ observable is suitable for measurement purposes, its lack of Lorentz covariance means it is not a suitable basis for a conserved 4-current density at other times. On the other hand, the Lorentz covariant expression in Eq. (2) seems suitable to be identified as the conserved 4-current density between measurements, but its time component in Eq. (3) is not suitable as a probability density for measurement outcomes.

A resolution of this dilemma will now be formulated over the next three sections. As a first step, the discussion in Sec. 4 examines the properties of expression (2) in more detail and highlights its domain of its applicability.

---

[5] $\mathrm{j}^0$ is then usually interpreted as charge density instead.

[6] e.g., Schweber [12], Sec. 3c.

## 4. Provisional suitability of the Klein-Gordon 4-current density

The standard Klein-Gordon 4-current density given in Eq. (2) has certain unusual properties which should be mentioned here. Its zeroth component alternates between positive and negative values, which means that the 4-current density lines in spacetime must have sections which point backwards in time, in addition to the usual forwards-in-time parts[7]. Also, these lines curve continuously and smoothly which means the current density 4-vector must pass through spacelike directions as well. Although perhaps surprising, this behaviour will not pose any conflict with experiment if it can be restricted to times between measurements and is not actually observed. Hence expression (2) remains viable subject to this restriction, which will require it to be interpreted as an average in Secs. 5 and 6.

This expression is actually the only one available from which to build a particle model because it is the only one ensured by the standard Klein-Gordon formalism to satisfy the continuity equation and thereby provide local conservation of probability[8]. It will therefore be tentatively adopted under the proviso that the next measurement performed is not one of position $\mathbf{x}$, thereby side-stepping the negative probability issue. In terms of unorthodox world lines remaining hidden, any observable (including position $\mathbf{q}$) for this measurement would be satisfactory instead. It will now be demonstrated how this necessary but, at first sight, seemingly artificial restriction can actually be shown to be natural and expected, as explained below.

## 5. Conditional 4-current density

The situation is clarified by following the procedure of previous authors [7-11] and introducing a quantity that is conditional on the final state. The relevant quantity in this case is the conditional 4-current density $j^{\alpha}\left(x\middle|i,f\right)$ for position $\mathbf{x}$ at time t given not only that the initial state was i but also that the subsequent measurement result is f. This quantity will be related to the usual 4-current density $j^{\alpha}\left(x\middle|i\right)$ in Eq. (2) and the probability density $\rho\left(f\right)$ in Eq. (4) via the following familiar relationship for conditional probabilities[9]:

$$j^{\alpha}\left(x\middle|i\right)=\int j^{\alpha}\left(x\middle|i,f\right)\rho\left(f\right)df \tag{6}$$

Inserting Eqs. (2) and (4) into Eq. (6) then yields the following more specific result:

$$\frac{i}{2m}\langle i|x\rangle\overset{\leftrightarrow}{\partial^{\alpha}}\langle x|i\rangle=\int j^{\alpha}\left(x\middle|i,f\right)\left|\langle f|i\rangle\right|^{2}df \tag{7}$$

---

[7] See, e.g., [14], particularly Fig. 1.

[8] Conservation laws are typically related to, and derived from, symmetries via Noether's theorem and Eq. (2) is the only conserved 4-current density provided by this theorem for the Klein-Gordon case.

[9] The integral here may actually be over multiple dimensions. For example, if the general observable represented here by f is chosen to be Newton and Wigner's $\mathbf{q}$ position in particular, the integral is then seen to be of the form $\iiint d^{3}q$.

To complete the present argument, there is a need to identify a viable expression for $j^{\alpha}(x|i,f)$. In addition to satisfying Eq. (7), this expression will need to be real, normalised and satisfy the continuity equation. Under these restrictions, the obvious choice is:

$$j^{\alpha}(x|i,f) = \frac{i}{2m}\mathrm{Re}\frac{\langle f|x\rangle \overset{\leftrightarrow}{\partial^{\alpha}} \langle x|i\rangle}{\langle f|i\rangle} \qquad (8)$$

which can be readily checked to have the required properties. Expression (8) then gives the conditional 4-current density at x given both the initial state i and the final measurement result f. For the purposes of the present argument, this expression will be assumed to represent the distribution of particle trajectories in an ensemble at intermediate times.

Eq. (8) has the form of a Bohmian guidance equation and carries the implication that the velocity of a particle depends not only on the state i but also on the type and outcome of a measurement to be conducted in the future, as will be touched on further in Sec. 7.

## 6. Automatic viability of the standard Klein-Gordon expression

The new expression (8) can now form the basis for successfully interpreting the standard Klein-Gordon 4-current density. Specifically, the standard expression (2) is seen to be the result of starting with the conditional expression (8) and then integrating via Eq. (6) over all the possible values for the unknown future result f. This procedure is based on the notion that the future state is not generally known and so must be averaged out. In this context, note that the structure of expression (8) automatically incorporates the required condition that the next measurement be of observable f and not of "non-observable" $\mathbf{x}$, as stipulated in Sec. 4. Now the important point to note here is that the standard Klein-Gordon expression in Eq. (2) will also be subject to this condition since it is simply defined here as the weighted average taken over the new expression (8). Stated explicitly, its definition automatically incorporates the fact that the next measurement is of observable f and not of "non-observable" $\mathbf{x}$. Now, as pointed out earlier, this is the necessary condition for the standard Klein-Gordon expression to be physically interpreted in terms of particle trajectories. Hence, introducing the new expression (8) allows the Klein-Gordon 4-current density to be reconciled with a particle ontology for quantum mechanics, with Eq. (2) now understood as representing the result of taking an average over f.

## 7. Properties of the proposed new 4-current density

The new conditional expression (8) is seen, as a result of the preceding considerations, to be fully viable for describing particle motion between measurements. It has some unusual properties, but none which raises any problem. Although it involves current lines which have both spacelike and backwards-in-time segments, these are already present in the standard Klein-Gordon 4-current density and, in any case, are hidden because they only occur at times between measurements (the actual outcomes being i and f). The important thing is that the probability density $|\langle f|i\rangle|^2$ for the next measurement result is positive, as required. Although the time component of expression (8) is not always positive, this is not important because its role is only to describe the direction of this 4-vector in spacetime and need not be used for predicting probabilities for experiments. Probabilities are only needed for measurements that are actually performed. Hence there is no requirement here to give a meaning to "negative probabilities", other than to take the sign as indicating a temporary reversal in time of the world line.

In any case, it should be remembered that this component is always positive in the local rest frame[10] of the 4-current density even though not always in our frame. Hence it could be interpreted locally as a probability density in the local rest frame.

From our 3-dimensional perspective, a consequence of the world line's switchback behaviour is that the particle actually becomes three particles temporarily, although invisibly. Nevertheless, the Klein-Gordon equation remains a single-particle equation in terms of the standard observations which can be performed.

The most novel feature of (8) is perhaps that it contains the future state $\left| f \right\rangle$, which means that the current $j^{\alpha}\left( x \middle| i,f \right)$ at x is being influenced retrocausally by the choice of measurement performed at a later time. Nevertheless it has already been shown elsewhere (see [11], Sec. 3 and [8], Sec. 2) that such an effect is unavoidable in any particle interpretation of quantum mechanics if Lorentz covariance is to be maintained in the many-particle case[11]. Hence adopting expression (8) as a correct description not only provides a viable interpretation of the Klein-Gordon 4-current density but is also consistent with the previous works cited.

## 8. Conclusion

A suitable interpretation of the Klein-Gordon 4-current density has been formulated within the context of a particle ontology for quantum mechanics. This has been achieved via the simple step of defining a quantity closely related to the Klein-Gordon expression, but dependent on the final state. This quantity shares the unusual features of the standard expression, but its structure automatically shields these features from experimental detection while remaining consistent with the predictions of quantum mechanics. In so doing, the new expression successfully explains the negative values encountered within the standard theory.

## Acknowledgement

The author wishes to thank Ken Wharton for his helpful feedback on this paper.

## Funding Acknowledgement

There was no funding.

---

[10] It is straightforward to extend the concept of a reference frame to superluminal velocities [15].

[11] The existence of retrocausal influences has been advocated by many other authors – see, e.g., [16] and references therein.

## Appendix

In going to a relativistic context (with, e.g., the Dirac or the Klein-Gordon equation) it is well known that position no longer satisfies the usual requirements for being an observable quantity. In the non-relativistic case it is easily shown that any two eigenstates of position are orthogonal:

$$
\begin{aligned}
\langle \mathbf{x} | \mathbf{x}' \rangle &= \int \langle \mathbf{x} | \mathbf{p} \rangle \langle \mathbf{p} | \mathbf{x}' \rangle \, d^3p \\
&= \int (2\pi)^{-3/2} \exp(i\mathbf{p}\cdot\mathbf{x}) \, (2\pi)^{-3/2} \exp(-i\mathbf{p}\cdot\mathbf{x}') \, d^3p \\
&= (2\pi)^{-3} \int \exp[i\mathbf{p}\cdot(\mathbf{x}-\mathbf{x}')] \, d^3p \\
&= \delta^3(\mathbf{x}-\mathbf{x}')
\end{aligned}
$$

This is not true, however, in the relativistic case where, using the notation $x \equiv (x^0, \mathbf{x})$ and $p \equiv (p^0, \mathbf{p})$, the corresponding result can be found via the following Lorentz invariant expression:

$$
\begin{aligned}
\langle x | x' \rangle &= \int \langle x | p \rangle \frac{m}{p^0} \langle p | x' \rangle \, d^3p \qquad [p^0 \equiv (\mathbf{p}\cdot\mathbf{p} + m^2)^{1/2}] \\
&= \int (2\pi)^{-3/2} \exp(-ip^\alpha x_\alpha) \frac{m}{p^0} (2\pi)^{-3/2} \exp(ip^\beta x'_\beta) \, d^3p \qquad (\alpha, \beta = 0,1,2,3) \\
&= (2\pi)^{-3} \int \frac{m}{p^0} \exp[-ip^\alpha (x_\alpha - x'_\alpha)] \, d^3p
\end{aligned}
$$

For equal times $x_0 = x'_0$ this reduces to:

$$
\begin{aligned}
\langle x | x' \rangle &= (2\pi)^{-3} \int \frac{m}{p^0} \exp[i\mathbf{p}\cdot(\mathbf{x}-\mathbf{x}')] \, d^3p \\
&\neq \delta^3(\mathbf{x}-\mathbf{x}')
\end{aligned}
$$

which shows that the position eigenstates are not orthogonal.